\documentclass[12pt,a4paper]{article}
\usepackage{graphicx}        
\usepackage{calc}
\newlength{\depthofsumsign}
\newcommand{\nsum}[1][1.4]{
	\mathop{%
		\raisebox
		{-#1\depthofsumsign+1\depthofsumsign}
		{\scalebox
			{#1}
			{$\displaystyle\sum$}%
		}
	}
}
\newlength{\depthofprodsign}
\newcommand{\nprod}[1][1.4]{
	\mathop{%
		\raisebox
		{-#1\depthofprodsign+1\depthofprodsign}
		{\scalebox
			{#1}
			{$\displaystyle\prod$}%
		}
	}
}
\title{Probability distribution connected to  stationary flow of substance in a 
	channel of network  containing finite number of arms}
\author{Roumen Borisov$^1$, Zlatinka I. Dimitrova$^2$, Nikolay K. Vitanov$^1$}
\date{$^1$Institute of Mechanics, Bulgarian Academy of Sciences, Acad. G. Bonchev Str., Bl. 4, 1113 Sofia, Bulgaria\\ 
$^2$"G. Nadjakov" Institute of Solid State Physics, Bulgarian Academy of Sciences, Blvd. Tzarigradsko Chaussee 72, 1784, Sofia, Bulgaria }
\begin{document}
\maketitle
\begin{abstract}
We discuss a channel consisting of nodes of a network and lines which 
connect these nodes and form ways for motion of a substance through the channel.  
 We study stationary flow of substance for channel which  arms 
contain finite number of nodes each and obtain probability distribution for  substance in arms of this channel. Finally we calculate  Shannon
	information measure for the case of stationary flow of substance in a simple 
	channel consisting of a single arm having just three nodes.
\end{abstract}
\section{Introduction}
In the course of time researchers began to study systems with increased complexity in different areas of sciences such as, for example,  natural sciences \cite{cs3} -\cite{boeck}, \cite{dim1}, \cite{dim2}  population dynamics \cite{dim3} - \cite{dim6}, \cite{ml}, \cite{sat},\cite{vit1},\cite{vit5} or social sciences \cite{cs2},\cite{cs4},\cite{cs5}. Large number of structures and processes in these complex systems can be modeled by  networks and differential equations \cite{chen1}, \cite{estrada}, \cite{fu}, \cite{ksv}, 
\cite{mv94} - \cite{v11b}, \cite{wu}. Our interest is concentrated on  motions of  substance through  channels of complicated  structures. These motions are  theoretically interesting and of significant practical 
importance  \cite{cs6}, \cite{cs7} as they are connected to transportation  
problems \cite{ff}, \cite{hani}, migration flows \cite{mf1}, \cite{mf3}, \cite{v1} or 
other kinds of flows in  networks \cite{ch1}. In this article
flow of substance in a channel which has have arbitrary  number of  arms. 
The  channel has special  nodes where split of an arm happens. 
More than one arm may arise by this split. The substance can move only in one 
direction along the channel. The substance can also move out of the channel -  "leakage" of substance.
\par 
The organization of text is as follows. In Sect. 2 we discuss a mathematical model 
for flow of substance in  studied channel for the case when each channel arm 
has finite number of nodes. We obtain probability distribution connected to  
amounts of substance in  nodes of  arms of this channel for the  case of stationary flow of substance. 
  In Sect. 3 we discuss information measures connected to the obtained probability distributions.

\section{Flow of substance in a channel with finite number of nodes in each arm}
The discussed model of motion of substance   is an
extension of  model discussed in \cite{v1}, \cite{sg}. The channel 
contains chains of nodes of the network  and the convention for numbering of nodes of  channel is as follows. We  assign  4 indexes to each node: ${\cal{V}}^{a,b}_{i,j}$. Lower indexes 
specify  position of  node in  current arm. $i$ is  associated with  current arm. 
$j$ is the  number of  node of the $i$-th arm. Upper indexes specify  the   
origin of  arm $i$.  Index $a$ denotes  number of  arm from which  
arm $i$ splits. Index $b$ denotes  number of  node of the arm $a$ where this split happens.  
Then ${\cal{V}}^{8,4}_{4,6}$ means the $6$-th node of  arm $4$ which splits at
node $4$ of  channel's arm $8$. The node where  arm $i$ begins is labeled as the 
$0$-th node of  $i$-th arm. This node is the next one after  splitting at  
node $b$ of  arm $a$. 
\par
The motion of substance in arms of studied channel is as follows.
Substance enters the channel from  external environment only through channel's main arm (labeled by $q=0$ 
below in the text). The substance can  move only in one direction in any arm: from nodes labeled by smaller number 
to nodes labeled by larger numbers. 
The nodes of each arm are connected by lines and each node is connected only
to  two neighboring nodes of the arm exclusive for  nodes where a
split of an arm happens. Last nodes can be connected to one or more additional nodes.
We assume that  substance can quit the channel and
can move to  environment - "leakage" of substance. As substance can enter the channel 
only through  $0$-th node of  main arm  then leakage is possible only in direction from  channel to environment and not in opposite direction. 
\par
We  consider each  arm to be an  array of of cells indexed in succession by non-negative integers.
We assume that an amount $x^{a_q,b_q}_q$ of some substance  is distributed among  
cells of the arm $q$  which splits at  node ($a_q,b_q$) of the network. This substance 
can move from one cell to next cell.  
Let $x_{q,i}^{a_q,b_q}$ be the amount of  substance in  $i$-th cell of  $q$-th arm of  channel. 
We consider first a channel containing  number of nodes $N_q+1$ in each of its  arms (the number of arms is $M$). For this case
$x_q^{a_q,b_q} = \nsum \limits_{i=0}^{N_q} x_{ q,i}^{a_q,b_q}$.
The fractions $y_{q,i}^{a_q,b_q} = x_{q,i}^{a_q,b_q}/x_{q}^{a_q,b_q}$ are considered as
probability values of distribution of a discrete random variable $\zeta$ in  corresponding arm of
channel:  $y_{q,i}^{a_q,b_q} = p_q^{a_q,b_q}(\zeta = i), \ i=0,1, \dots$.
The amount $x_{q,i}^{a_q,b_q}$ of substance in  $i$-th node of  $q$-th arm 
of  channel can change because of the following processes. First of all some amount $s^{a,b}_{q}$ of 
substance   enters  arm $q$ from  external environment through  $0$-th cell of  the arm. We consider two kinds of external environments. For the root of the channel (arm with label $q=0$):  substance $s_0^{0,0}$ enters the root through  environment of the channel. For the arms of the channel that are not  root ($q \ne 0$) : substance 
$s_q^{a,b}$ is part of  substance presented in  node $(a,b)$ of  parent arm. 
\par
The substance $s_q^{a,b}$ is presented only in   node $0$ of  $q$-th arm of  channel.  For  other nodes of  channel there is no substance that enters the node from  environment or from  other arms of the channel. Next
amount  $f_{q,i}^{a,b}$ from $x_{q,i}^{a,b}$ is transferred from the $i$-th cell into  $i+1$-th cell of  $q$-th arm and amount $g_{q,i}^{a,b}$ of  $x_{q,i}^{a,b}$  leaks out the $i$-th cell of $q$-th arm into   environment of the arm of the channel. This leakage can be of two kinds. First of all there can be leakage to the environment of the channel. This leads to loss of substance for the channel. In addition there can be a leakage to other arms of the channel which begin from  node $b$ of  arm $a$. This leakage is connected to the substance $s_q^{a,b}$ that enters  corresponding child arm of the channel which  splits from  node $b$ of  arm $a$.
\par
The process of motion of  substance is continuous in  time and  motion of substance among  
nodes of  $q$-th channel is  modeled mathematically by a system of ordinary differential 
equations which contains an equation for  $0$-th node, equations for  nodes $1,\dots, N_{q-1}$ 
and equation for  node $N_q$. This system is
$\frac{dx_{q,0}^{a,b}}{dt} = s_q^{a,b} - f_{q,0}^{a,b} - g_{q,0}^{a,b}$; 
$\frac{dx_{q,i}^{a,b}}{dt} = f_{q,i-1}^{a,b} -f_{q,i}^{a,b}  - g_{q,i}^{a,b}$,  $i=1,2,N_q-1$
$\frac{dx_{q,N_q}^{a,b}}{dt} = f_{q,N_q-1}^{a,b}  - g_{q,N_q}^{a,b}$.
We shall discuss below  the stationary regime of functioning of channel where
$dx_{i,q}^{a,b}/dt=0$, $i=0,1,N_q$. We mark the quantities for the stationary case with $^*$. Then from
equations above one obtains:
$f_{q,0}^{* a,b}  =  s_q^{* a,b}  - g_{q,0}^{* a,b}$; \   $f_{q,i}^{* a,b} = f_{q,i-1}^{* a,b}  - g_{q,i}^{* a,b}$;
\ $f_{q,N_q}^{* a,b} = f_{q,N_q-1}^{* a,b}  - g_{q,N_q}^{* a,b}$.
We  assume  the following forms of  amounts of  moving substances  in above equations
($\alpha_i, \beta_i, \gamma_i, \sigma$ are parameters):
$ s_0^{0,0} = \sigma_0 x_{0,0}^{0,0}  > 0 ; \ \
s_{q}^{a,b}= \delta_q x_{a,b}^{c,d}  \ \ (1 \geq \delta_q \geq 0); 
f_{q,i}^{a,b} = \alpha_{q,i}^{a,b}   x_{q,i}^{a,b}  \ \ (1 > \alpha_{q,i}^{a,b} >0)$; 
$g_{q,i}^{a,b} = \gamma_{q,i}^{*a,b} x_{q,i}^{a,b}; \ \ \ 1 \geq \gamma_{q,i}^{*a,b}\ge 0 \to \textrm{non-uniform leakage  in the nodes}$.
Indexes $c$ and $d$ above describe  parent arm (numbered by $c$) 
and  parent node of the arm $c$ (numbered by $d$) for the arm $q$. 
$\gamma_{q,i}^{*a,b} = \gamma_{q,i}^{a,b} + \nsum \limits_{p \in (q,i)}\delta_{p, q,i}^{a,b} $ 
describes the situation with  leakages in  cells where new arm arises and part of  substance flows to this new arm.
$\gamma_{q,i}^{a,b}$ is the leakage to  environment from  $i$-th node of  $q$-th arm.
Notation $p \in (q,i)$ in the sum  denotes all the arms which arise from  node $(q,i)$.
The terms in the sum describe leakages because of outflow of substance from
parent arm to child arms.
The model system of equations for  node $0$
and for  nodes $1,\dots, N_{q-1}$ of  $q$-th arm of the channel is
\begin{eqnarray} \label{eqa1}
\frac{dx_{q,0}^{a,b}}{dt}&=&s_q^{a,b}-\alpha_{q,0}^{a,b} x_{q,0}^{a,b} - \gamma_{q,0}^{*a,b} x_{q,0}^{a,b} , \nonumber \\
\frac{dx_{q,i}^{a,b}}{dt}&=& \alpha_{q,i-1}^{a,b} x_{q,i-1}^{a,b}-(\alpha_{q,i}^{a,b} +\gamma_{q,i}^{*a,b})x_{q,i}^{a,b};\ \ i=1,2,\dots, N_q -1 . 
\end{eqnarray}
For node $N_q$ of  $q$-th arm  there is no outflow to  next mode of the arm 
(as the node $N_q$ is the last node of  $q$-th arm) and the equation for  motion of substance for this node is
\begin{eqnarray}\label{eqa3}
\frac{dx_{q,N}^{a,b}}{dt}&=&\alpha_{q,N_q-1}^{a,b} x_{q,N_q-1}^{a,b}-\gamma_{q,N_q}^{*a,b}x_{q,N_q}^{a,b}.
\end{eqnarray}
We discuss   stationary motion of  substance through  arms of studied channel.
Then $dx^{a,b}_{q,0}/dt=0$ in the first of Eqs. (\ref{eqa1}) and 
\begin{equation}\label{eqa4}
x_{q,0}^{*a,b}=\frac{s_q^{a,b}}{\alpha_{0,q}^{a,b}+{\gamma_{0,q}^{*,a,b}}} \ \ .
\end{equation}
For  root of the channel (arm $0$) we substitute $s_0^{0,0}$  in (\ref{eqa4}) and obtain that
$x_{0,0}^{0,0}$ is a free parameter and in addition 
$\sigma_0 = \alpha_{0,0}^{0,0} + \gamma_{0,0}^{*,0,0}.$
For  arm $r$  which arises from node $m$ of  arm $q$,  $dx_{r,0}^{q,m}/dt=0$ and thus 
we obtain
\begin{equation}\label{eqa5}
x_{r,0}^{*q,m} = \frac{\delta_{r,q,m}^{a,b} }{\alpha_{r,0}^{q,m} + \gamma_{r,0}^{q,m}} x_{q,m}^{*c,d}.
\end{equation}
For $x_{q,i}^{*a,b}$ one obtains the relationship (just set $dx_{q,i}^{a,b}/dt = 0$ in  second of Eqs. (\ref{eqa1}))
\begin{equation}\label{eqa6}
x_{q,i}^{*a,b} = \frac{\alpha_{q,i-1}^{a,b}}{\alpha_{q,i}^{a,b} + \gamma_{q,i}^{*a,b}} x_{q,i-1}^{*a,b}, \ i=1,2,\dots, N_q-1;
\end{equation}
In order to calculate  $x_{q,N_q}^{*a,b}$ we have to use (\ref{eqa3}). The result is
\begin{equation}\label{eqa7}
x_{q,N_q}^{*a,b} = \frac{\alpha_{q,N_q-1}^{a,b}}{\gamma_{q,N_q}^{*a,b}} x_{q,N_q-1}^{*a,b}. 
\end{equation}
From (\ref{eqa6}) we obtain
\begin{equation}\label{eqa8}
x_{q,i}^{*a,b} = \frac{\nprod[2] \limits_{j=0}^{i-1} 
	\alpha_{q,i-j-1}^{a,b}}{
	\nprod[2] \limits_{j=0}^{i-1} \bigg [ \alpha_{q,i-j}^{a,b}  + 
	\gamma^{*a,b}_{q,i-j}\bigg ]} x_{q,0}^{*a,b}, \ \ 
i=1,\dots,N_q-1.
\end{equation}
From Eq.(\ref{eqa7}) one obtains
\begin{equation}\label{eqa9}
x_{q,N_q}^{*a,b} = \frac{\alpha_{q,N_q-1}^{a,b}}{\gamma_{q,N_q}^{*a,b}} 
\frac{\nprod[2] \limits_{j=0}^{N_q-2}  \alpha_{q,N_q-j-2}^{a,b}}{
	\nprod[2] \limits_{j=0}^{N_q-2} \bigg[ \alpha_{q,N_q-j-1}^{a,b}  + \gamma^{*a,b}_{q,N_q-j-1} \bigg]} x_{q,0}^{*a,b}.
\end{equation}
The total amount of the substance in  $q$-th arm of the channel is
{\small
\begin{eqnarray}\label{eqa10}
x_q^{*,a,b} =  x_{q,0}^{*a,b} +   x_{q,i}^{*,a,b} + x_{q,N}^{*,a,b} =
x_{q,0}^{*a,b} \times \nonumber \\
\left ( 1 + 
\nsum[4] \limits_{i=1}^{N_q-1} \frac{\nprod[2] \limits_{j=0}^{i-1} \alpha_{q,i-j-1}^{a,b}}{
	\nprod[2] \limits_{j=0}^{i-1} \bigg [ \alpha_{q,i-j}^{a,b} + 
	\gamma^{*a,b}_{q,i-j}\bigg ]} \right . + \left. \frac{\alpha_{q,N_q-1}^{a,b}}{\gamma_{q,N_q}^{*a,b}} 
\frac{\nprod[2] \limits_{j=0}^{N_q-2} \alpha_{q,N_q-j-2}^{a,b}}{
	\nprod[2] \limits_{j=0}^{N_q-2} \bigg[ \alpha_{q,N_q-j-1}^{a,b}  + 
	\gamma^{*a,b}_{q,N_q-j-1} \bigg]}  \right ) \nonumber \\	
\end{eqnarray}
}
The distribution of substance in nodes of  $q$-th arm of the channel is
{\small 
\begin{eqnarray}\label{eqa11}
y_{q,0}^{*a,b}= 1 \Bigg / \nonumber \\
 \left ( 1 + 
\nsum[4] \limits_{i=1}^{N_q-1} \frac{\nprod[2] \limits_{j=0}^{i-1} \alpha_{q,i-j-1}^{a,b}}{
	\nprod[2] \limits_{j=0}^{i-1} \bigg [ \alpha_{q,i-j}^{a,b}  + 
	\gamma^{*a,b}_{q,i-j}\bigg ]} \right . + 
\left. \frac{\alpha_{q,N_q-1}^{a,b}}{\gamma_{q,N_q}^{*a,b}} 
\frac{\nprod[2] \limits_{j=0}^{N_q-2}  \alpha_{q,N_q-j-2}^{a,b}}{
	\nprod[2] \limits_{j=0}^{N_q-2} \bigg[ \alpha_{q,N_q-j-1}^{a,b}  + 
	\gamma^{*a,b}_{q,N_q-j-1} \bigg]}  \right ) \nonumber \\
\end{eqnarray}
}
{\small
\begin{eqnarray}\label{eqa12}
y_{q,i}^{*a,b}= \left( \frac{\nprod[2] \limits_{j=0}^{i-1} \alpha_{q,i-j-1}^{a,b}}{
	\nprod[2] \limits_{j=0}^{i-1} \bigg [ \alpha_{q,i-j}^{a,b}  + 
	\gamma^{*a,b}_{q,i-j}\bigg ]} \right) \Bigg / \nonumber \\ 
	\left ( 1 +  
\nsum[4] \limits_{i=1}^{N_q-1} \frac{\nprod[2] \limits_{j=0}^{i-1} \alpha_{q,i-j-1}^{a,b}}{
	\nprod[2] \limits_{j=0}^{i-1} \bigg [ \alpha_{q,i-j}^{a,b}  + 
	\gamma^{*a,b}_{q,i-j}\bigg ]}  + \right. 
\left. \frac{\alpha_{q,N_q-1}^{a,b}}{\gamma_{q,N_q}^{*a,b}} 
\frac{\nprod[2] \limits_{j=0}^{N_q-2}  \alpha_{q,N_q-j-2}^{a,b}}{
	\nprod[2] \limits_{j=0}^{N_q-2} \bigg[ \alpha_{q,N_q-j-1}^{a,b} + 
	\gamma^{*a,b}_{q,N_q-j-1} \bigg]}  \right ) \nonumber \\ i=1,\dots,N_q-1\nonumber \\
\end{eqnarray}
}
{\small
\begin{eqnarray}\label{eqa13}
y_{q,N_q}^{*a,b}= \left( \frac{\alpha_{q,N_q-1}^{a,b}}{\gamma_{q,N_q}^{*a,b}} 
\frac{\nprod[2] \limits_{j=0}^{N_q-2}  \alpha_{q,N_q-j-2}^{a,b}}{
	\nprod[2] \limits_{j=0}^{N_q-2} \bigg[ \alpha_{q,N_q-j-1}^{a,b}  + \gamma^{*a,b}_{q,N_q-j-1} \bigg]} \right) \Bigg / \nonumber \\
\left ( 1 +  
\nsum[4] \limits_{i=1}^{N_q-1} \frac{\nprod[2] \limits_{j=0}^{i-1} \alpha_{q,i-j-1}^{a,b}}{
	\nprod[2] \limits_{j=0}^{i-1} \bigg [ \alpha_{q,i-j}^{a,b} + 
	\gamma^{*a,b}_{q,i-j}\bigg ]}  + \right. \left. \frac{\alpha_{q,N_q-1}^{a,b}}{\gamma_{q,N_q}^{*a,b}} 
\frac{\nprod[2] \limits_{j=0}^{N_q-2}  \alpha_{q,N_q-j-2}^{a,b}}{
	\nprod[2] \limits_{j=0}^{N_q-2} \bigg[ \alpha_{q,N_q-j-1}^{a,b}  + 
	\gamma^{*a,b}_{q,N_q-j-1} \bigg]}  \right ) \nonumber \\
\end{eqnarray}
}
\section{Information measures for studied channel} 
We have discussed  probability distributions corresponding to  a channel with
single arm  in \cite{bv2}, \cite{vkx1} - \cite{bv1}. Particular cases of these distributions 
are Waring distribution, Zipf distribution, Yule-Simon distribution, Binomial distribution, etc. 
We can study also other kinds of distributions on the basis of the model discussed
in this text. One example is  probability distribution
of the substance in a part of the studied network which contains several channels for motion of substance. 
We can calculate various characteristics of the distributions obtained above and let us
consider an information problem connected to the flow of substance  in studied 
channel.  Each node of the channel is numbered and we  consider 
the nodes as  letters of an alphabet. Some kind of an event happens in any of  nodes 
of the channel and let  probability of happening of this event be proportional to  amount of substance in corresponding node. 
Thus the probability of happening of  event in a node 
of  channel will be equal to the  probability from corresponding probability distribution obtained above in the text. 
The channel (the source) will generate events with corresponding probability and we can calculate measure of information and   
Shannon measure of information for these sequences as  probability distribution is known.
The information measure connected to an event with probability $p$ is
$I(p) = - \log(p)$,
and  Shannon information measure (average information we get from a symbol in a stream) 
connected to  probability distribution $P=(p_0,\dots,p_N)$ is
$H (P) = - \sum \limits_{i=0}^N p_i \log (p_i)$.
Let us consider  distribution $P^*$ of substance in the $q$-th arm of the channel given by (\ref{eqa11}) - (\ref{eqa13}).
The information connected to  event with probability $p_i$ from  $i$-th node of this arm  is
{\small
\begin{eqnarray}\label{s3}
I(p_i)=  -\log \left( \frac{\nprod[2] \limits_{j=0}^{i-1}  \alpha_{q,i-j-1}^{a,b}}{
	\nprod[2] \limits_{j=0}^{i-1} \bigg [ \alpha_{q,i-j}^{a,b}  
	\gamma^{*a,b}_{q,i-j}\bigg ]} \right) +	\log \left ( 1 +  
\nsum[4] \limits_{i=1}^{N_q-1} \frac{\nprod[2] \limits_{j=0}^{i-1} \alpha_{q,i-j-1}^{a,b}}{
	\nprod[2] \limits_{j=0}^{i-1} \bigg [ \alpha_{q,i-j}^{a,b}  + 
	\gamma^{*a,b}_{q,i-j}\bigg ]}  + \right. \nonumber \\
\left. \frac{\alpha_{q,N_q-1}^{a,b}}{\gamma_{q,N_q}^{*a,b}} 
\frac{\nprod[2] \limits_{j=0}^{N_q-2}  \alpha_{q,N_q-j-2}^{a,b}}{
	\nprod[2] \limits_{j=0}^{N_q-2} \bigg[ \alpha_{q,N_q-j-1}^{a,b}  + 
	\gamma^{*a,b}_{q,N_q-j-1} \bigg]}  \right ), \ 
i=1,\dots,N_q-1 \nonumber\\ 
\end{eqnarray}
}
For the $0$-th node and for the $N_q$-th node we obtain the following relationships for the information
measures
{\small
\begin{eqnarray}\label{s3a}
I(p_0)= 
\log  \left ( 1 + 
\nsum[4] \limits_{i=1}^{N_q-1} \frac{\nprod[2] \limits_{j=0}^{i-1} \alpha_{q,i-j-1}^{a,b}}{
	\nprod[2] \limits_{j=0}^{i-1} \bigg [ \alpha_{q,i-j}^{a,b}  + 
	\gamma^{*a,b}_{q,i-j}\bigg ]} \right . + \left. \frac{\alpha_{q,N_q-1}^{a,b}}{\gamma_{q,N_q}^{*a,b}} 
\frac{\nprod[2] \limits_{j=0}^{N_q-2}  \alpha_{q,N_q-j-2}^{a,b}}{
	\nprod[2] \limits_{j=0}^{N_q-2} \bigg[ \alpha_{q,N_q-j-1}^{a,b}  
	\gamma^{*a,b}_{q,N_q-j-1} \bigg]}  \right )  \nonumber \\ 
I(p_N)= 
-\log  \left( \frac{\alpha_{q,N_q-1}^{a,b}}{\gamma_{q,N_q}^{*a,b}} 
\frac{\nprod[2] \limits_{j=0}^{N_q-2} \alpha_{q,N_q-j-2}^{a,b}}{
	\nprod[2] \limits_{j=0}^{N_q-2} \bigg[ \alpha_{q,N_q-j-1}^{a,b}  + \gamma^{*a,b}_{q,N_q-j-1} \bigg]} \right) + \nonumber \\
	\log \left ( 1 +  
\nsum[4] \limits_{i=1}^{N_q-1} \frac{\nprod[2] \limits_{j=0}^{i-1} \alpha_{q,i-j-1}^{a,b}}{
	\nprod[2] \limits_{j=0}^{i-1} \bigg [ \alpha_{q,i-j}^{a,b}  + 
	\gamma^{*a,b}_{q,i-j}\bigg ]}  + \right.
\left. \frac{\alpha_{q,N_q-1}^{a,b}}{\gamma_{q,N_q}^{*a,b}} 
\frac{\nprod[2] \limits_{j=0}^{N_q-2}  \alpha_{q,N_q-j-2}^{a,b}}{
	\nprod[2] \limits_{j=0}^{N_q-2} \bigg[ \alpha_{q,N_q-j-1}^{a,b} + 
	\gamma^{*a,b}_{q,N_q-j-1} \bigg]}  \right ). 
\end{eqnarray}
}
\par
The general expression for the Shannon information measure is very long and we will not 
write it here. Instead of this we  present 
the Shannon information measure for a simple case of a channel arm that has just three nodes.  
We shall omit the indices $a$, $b$ and $q$. For discussed case $N_q=N=2$ as  labels of  
nodes of the arm are $0$, $1$ and $2$.  The parameters below account for  following processes in the studied arm.
$\alpha_0$ ($0<\alpha_0<1$) accounts for flow between first and second node; $\alpha_1$ ($0<\alpha_1<1$) accounts 
for flow between second and third node; $\beta_1$ ($0<\beta_1 < 1-\alpha_1$) accounts for preference for the 
third node; $\gamma_1^*$ ($0 \le \gamma_1^*<1-\alpha_1-\beta_1$) accounts for leakage from the second node; 
$\gamma_2^*$ ($0<\gamma_2^*\le 1$) accounts for leakage from the third node.
The Shannon information measure is
{\small
\begin{eqnarray}\label{s7}
H &=& \frac{1}{ \left [ 1 +\frac{\alpha_0}{\alpha_1+\beta_1+\gamma_1^*} \left(1+ \frac{\alpha_1+\beta_1}{\gamma_2^*} \right) \right]}
\log  \left [ 1 +\frac{\alpha_0}{\alpha_1+\beta_1+\gamma_1^*} \left(1+ \frac{\alpha_1+\beta_1}{\gamma_2^*} \right) \right] + \nonumber \\
&& \frac{ \left( \frac{\alpha_0}{\alpha_1+\beta_1+\gamma_1^*}  \right)}{ \left [ 1 +\frac{\alpha_0}{\alpha_1+\beta_1+\gamma_1^*} \left(1+ \frac{\alpha_1+\beta_1}{\gamma_2^*} \right) \right]} \left\{\log  \left [ 1 +\frac{\alpha_0}{\alpha_1+\beta_1+\gamma_1^*} \left(1+ \frac{\alpha_1+\beta_1}{\gamma_2^*} \right) \right] - \right. \nonumber \\
&& \left. \log \left( \frac{\alpha_0}{\alpha_1+\beta_1+\gamma_1^*}  \right)  \right \} + \frac{ \left(\frac{\alpha_1+\beta_1}{\gamma_2^*} \frac{\alpha_0}{\alpha_1+\beta_1+\gamma_1^*}  \right)}{ \left [ 1 +\frac{\alpha_0}{\alpha_1+\beta_1+\gamma_1^*} \left(1+  \frac{\alpha_1+\beta_1}{\gamma_2^*} \right) \right]}
\left \{   \log  \left [ 1 + \frac{\alpha_0}{\alpha_1+\beta_1+\gamma_1^*} \times \right. \right.  \nonumber \\
&& \left. \left.  \left(1+\frac{\alpha_1+\beta_1}{\gamma_2^*} \right) \right] -
\log \left(\frac{\alpha_1+\beta_1}{\gamma_2^*} \frac{\alpha_0}{\alpha_1+\beta_1+\gamma_1^*}  \right)  \right\}.
\end{eqnarray}
}
The studies show that $H$ can have a maximum value for some value of the parameters of the problem. This
means that  average information we can obtain from an events happening in the channel can have local 
maxima for some values of  problem parameters.

\end{document}